\documentclass[aps,prl,amsmath,amssymb,twocolumn,superscriptaddress,showpacs]{revtex4-1}
\usepackage{amsmath,amssymb}
\usepackage{graphicx}	

\usepackage[usenames,dvipsnames]{color}
\usepackage[colorlinks=true, citecolor=blue, linkcolor=blue, urlcolor=blue]{hyperref}

\def\be{\begin{eqnarray}}   
\def\ee{\end{eqnarray}}

\begin{document}

\author{S.~A.~Sato}
\email{shunsuke.sato@mpsd.mpg.de}
\affiliation 
{Max Planck Institute for the Structure and Dynamics of Matter, Luruper Chaussee 149, 22761 Hamburg, Germany}

\author{J.~W.~McIver}
\affiliation 
{Max Planck Institute for the Structure and Dynamics of Matter, Luruper Chaussee 149, 22761 Hamburg, Germany}

\author{M.~Nuske}
\affiliation 
{Zentrum f\"ur Optische Quantentechnologien and Institut f\"ur Laserphysik,
Universit\"at Hamburg, Luruper Chaussee 149, 22761 Hamburg, Germany}

\author{P.~Tang}
\affiliation 
{Max Planck Institute for the Structure and Dynamics of Matter, Luruper Chaussee 149, 22761 Hamburg, Germany}

\author{G.~Jotzu}
\affiliation 
{Max Planck Institute for the Structure and Dynamics of Matter, Luruper Chaussee 149, 22761 Hamburg, Germany}

\author{B.~Schulte}
\affiliation 
{Max Planck Institute for the Structure and Dynamics of Matter, Luruper Chaussee 149, 22761 Hamburg, Germany}

\author{H.~H\"ubener}
\affiliation 
{Max Planck Institute for the Structure and Dynamics of Matter, Luruper Chaussee 149, 22761 Hamburg, Germany}

\author{U.~De~Giovannini}
\affiliation 
{Max Planck Institute for the Structure and Dynamics of Matter, Luruper Chaussee 149, 22761 Hamburg, Germany}

\author{L.~Mathey}
\affiliation 
{Zentrum f\"ur Optische Quantentechnologien and Institut f\"ur Laserphysik,
Universit\"at Hamburg, Luruper Chaussee 149, 22761 Hamburg, Germany}
\affiliation
{The Hamburg Centre for Ultrafast Imaging, University of Hamburg,
Luruper Chaussee 149, 22761 Hamburg, Germany}

\author{M.~A.~Sentef}
\affiliation 
{Max Planck Institute for the Structure and Dynamics of Matter, Luruper Chaussee 149, 22761 Hamburg, Germany}

\author{A.~Cavalleri}
\affiliation 
{Max Planck Institute for the Structure and Dynamics of Matter, Luruper Chaussee 149, 22761 Hamburg, Germany}

\author{A.~Rubio}
\email{angel.rubio@mpsd.mpg.de}
\affiliation 
{Max Planck Institute for the Structure and Dynamics of Matter, Luruper Chaussee 149, 22761 Hamburg, Germany}
\affiliation 
{Center for Computational Quantum Physics (CCQ), Flatiron Institute, 162 Fifth Avenue, New York, NY
10010, USA}

\title{Microscopic theory for the light-induced anomalous Hall effect in graphene
}

\begin{abstract}

We employ a quantum Liouville equation with relaxation to model the recently observed anomalous Hall effect in graphene irradiated by an ultrafast pulse of circularly polarized light. In the weak-field regime, we demonstrate that the Hall effect originates from an asymmetric population of photocarriers in the Dirac bands. By contrast, in the strong-field regime, the system is driven into a non-equilibrium steady state that is well-described by topologically non-trivial Floquet-Bloch bands. Here, the anomalous Hall current originates from the combination of a population imbalance in these dressed bands together with a smaller anomalous velocity contribution arising from their Berry curvature. This robust and general finding enables the simulation of electrical transport from light-induced Floquet-Bloch bands in an experimentally relevant parameter regime and creates a pathway to designing ultrafast quantum devices with Floquet-engineered transport properties.

\end{abstract}

\maketitle

Optical control of functional materials has emerged as an important research front bridging condensed matter physics
\cite{Basov2017} and ultrafast spectroscopy \cite{Krausz2014}.
Many noteworthy phenomena have been discovered in optically-driven quantum solids, including light-induced superconductivity \cite{Fausti189,Mitrano2016},
various types of photo-initiated insulator-metal transitions
\cite{Fiebig1925,PhysRevLett.87.237401,Rini:05,Liu2012},
light control of microscopic interactions like electron-phonon coupling
\cite{PhysRevB.95.024304,Kennes2017,PhysRevB.95.205111},
and theoretically predicted Floquet-topological phases of matter 
\cite{PhysRevB.79.081406,PhysRevB.84.235108,Lindner2011,Sentef2015,Huebener2017}.
Floquet-topological phases in particular have stimulated much interest 
but direct evidence of electron-photon Floquet-dressed states in solids 
is scarce to this date
\cite{Wang453,Mahmood2016}, in contrast to the field of artificial lattices
\cite{Struck996,PhysRevLett.107.255301,PhysRevLett.111.185302,Jotzu2014,Flaschner1091,nager2018parametric,PhysRevLett.81.5093,PhysRevLett.96.243901,PhysRevLett.99.220403,Rechtsman2013,AIDELSBURGER2018394,PhysRevLett.115.073002,Parker2013,Kennedy2015,Aidelsburger2014,ozawa2018topological,asteria2018measuring}.

Recently a light-induced anomalous Hall effect was observed in graphene using ultrafast transport techniques \cite{2018McIver}. 
A key challenge for the interpretation of the reported effects lies in the competition between Floquet engineering of Hamiltonians versus the role of electronic population effects. For the case of laser-driven graphene, the latter are particularly important as the pump laser is generically resonant with electronic excitations. 
Here we provide a theoretical framework within which this class of experiments \cite{2018McIver} can be interpreted.

A graphene lattice subjected to circular driving has been studied 
theoretically in a variety of frameworks \cite{PhysRevB.79.081406,PhysRevB.84.235108,Lindner2011,Bukov2015,PhysRevLett.113.266801,Jotzu2014,PhysRevB.90.115423,PhysRevB.93.144307,PhysRevB.90.195429,PhysRevB.91.155422,Sentef2015,PhysRevLett.113.236803}.
We focus here on the low-frequency driving regime and find that the driven-dissipative dynamics together with the applied bias field plays a crucial role in understanding the transport properties of the Floquet-engineered state.
Our real-time simulations contain both 
the population imbalance of excited photocarriers in the Dirac cone of graphene as well as 
the Floquet-topological Berry curvature of photon-dressed bands. 
We find that population effects play an important role under the low-frequency driving used in the experiments 
in both weak and strong driving limits. 
In the weak-driving regime, light-induced Hall transport originates mainly from population imbalance of photocarriers in the bare bands.  By contrast, in the strong field regime clear topological-Floquet states are formed, and the light-induced Hall effect is characterized by a population imbalance of these Floquet bands that outweighs, however, the Floquet-Berry  curvature  contribution which is  predicted  to  dominate  in the high-frequency regime \cite{PhysRevB.79.081406,PhysRevLett.113.266801,PhysRevB.90.195429}. Our results demonstrate that Floquet-engineering in solids is a reality, even with significant dissipation. These findings 
are in good agreement with the experimental results \cite{2018McIver} and provide a microscopic interpretation of the observed light-induced anomalous Hall effect.

To model the electron dynamics in graphene under electromagnetic fields,
we employ a quantum Liouville equation for the one-particle reduced
density matrix $\rho_{\boldsymbol{K}(t)}$,
\be
\frac{d}{dt}\rho_{\boldsymbol{K}(t)}(t) &=& 
\frac{\left [ H_{\boldsymbol{K}(t)},\rho_{\boldsymbol{K}(t)}(t) \right ]}{i\hbar} 
+ \hat D_{\boldsymbol{K}(t)} \rho_{\boldsymbol{K}(t)}(t), \label{eq:liouville}
\ee
with phenomenological relaxation $\hat D_{\boldsymbol K(t)}$ and
the Dirac Hamiltonian, $H_{\boldsymbol{K}(t)} = \hbar v_F \tau_z \sigma_x K_x(t) 
+ \hbar v_F \sigma_y K_y(t)$,
where $v_F$ is the Fermi velocity, $\sigma_{x/y}$ are the Pauli matrices, and
$\boldsymbol{K}(t)=\boldsymbol{k}+e\boldsymbol{A}(t)/\hbar c$ is the lattice momentum coupled to an external vector potential
$\boldsymbol{A}(t)$. The linear Dirac-cone approximation is justified in the present context of low-frequency driving and moderate deviations from half-filled bands
with a chemical potential shift $\mu$ \cite{2018McIver}.
The different chirality of the Dirac fermions at the $K$ and $K'$
points is given by $\tau_z = \pm 1$.
The phenomenological dissipation $\hat D_{\boldsymbol{K}(t)}$ 
based on the relaxation time approximation is added to account for relaxation and dephasing effects
\cite{PhysRevLett.73.902, 2018satoSI}.
We note that effects of dissipatively coupled Floquet-topological systems have been discussed previously
\cite{PhysRevB.90.195429,PhysRevB.91.155422,PhysRevX.5.041050,PhysRevB.93.144307},
but not in the concrete context of the present work.
Here we set the relaxation time $T_1$ to $100$~fs, and the dephasing time
$T_2$ to $20$~fs. However, the qualitative behavior of the light-induced Hall effect
does not strongly depend on the choice of $T_1$ and $T_2$.

\begin{figure}[htbp]
  \includegraphics[width=0.9\columnwidth]{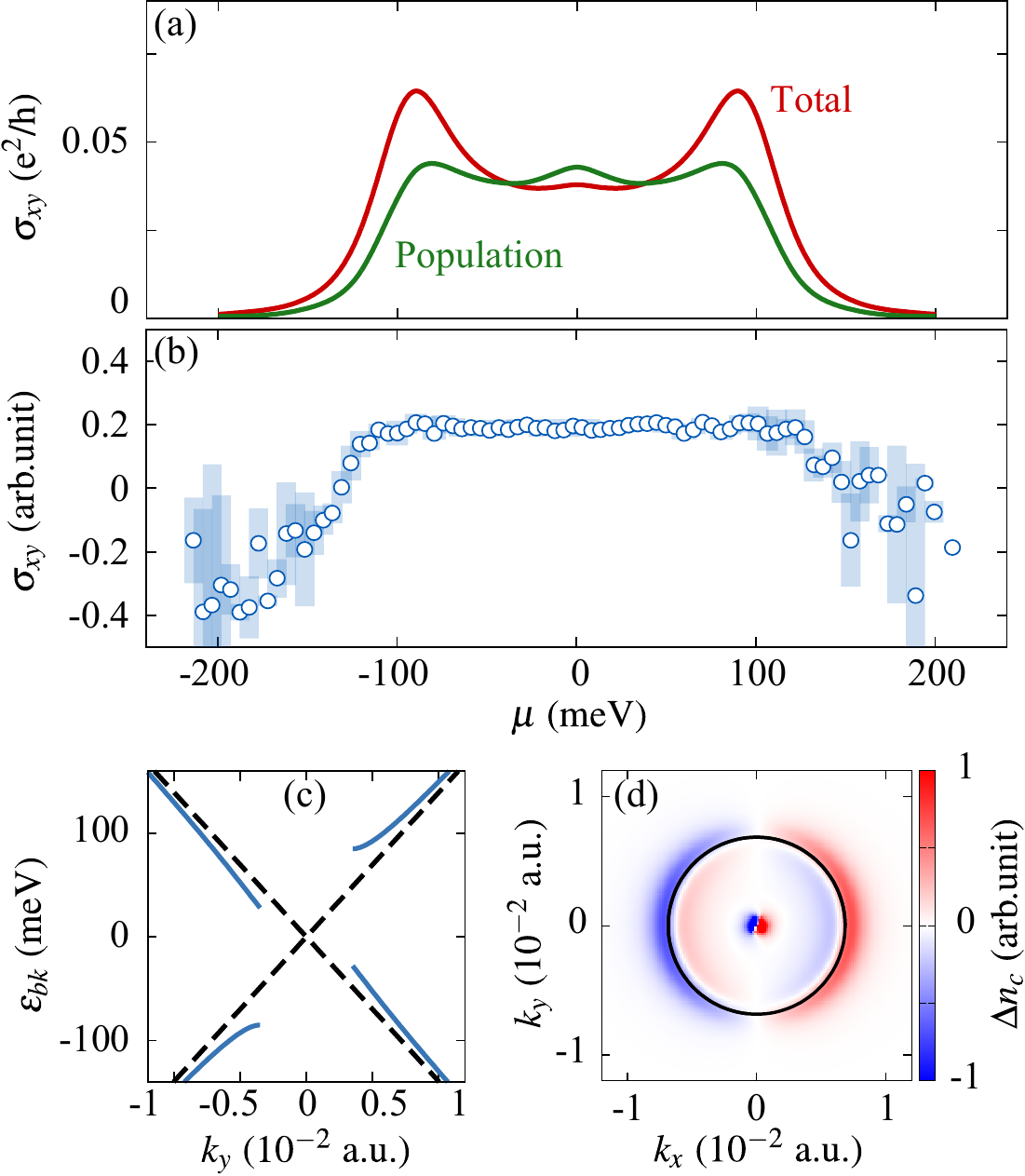}
\caption{\label{fig:fig1}
Light-induced Hall conductivity in the weak-field regime,
$E_{\text{\text{\text{MIR}}}}=1$~MV/m:
(a)~Theoretical Hall conductivity $\sigma_{xy}$ 
as a function of $\mu$.
The full simulation result (red) and the population contribution (green)
are shown. (b)~The corresponding experimental Hall conductivity as shown in Ref.~\cite{2018McIver}.
(c)~Electronic structure of the Dirac Hamiltonian. Black-dashed lines show
the original Dirac bands, while blue-solid lines show the tilted Dirac bands under a source-drain field with strength $0.2$~a.u.
(d)~Pump dichroism of conduction-band populations under source-drain bias along the $y$-direction.
}
\end{figure}

To evaluate the Hall conductivity $\sigma_{xy}$,
we apply a weak static source-drain electric field along the $y$-axis of the model
and compute the transverse current along the $x$-axis 
under the presence of a circular laser pulse. 
Both the static electric field and the laser pulse are included in
the model via the vector potential $\boldsymbol{A}(t)$.
Following the same analysis as in Ref.~\onlinecite{2018McIver},
we define the Hall current as the difference of the transverse current
induced by right- and left-handed circular laser fields (pump dichroism) \cite{2018satoSI}. 
To match experimental values we set the wavelength of the circular laser pulse
to $\lambda=6.5$~$\mu$m, which corresponds to the mean photon energy
of $\hbar \omega_{\text{MIR}}\approx190$~meV. The pulse duration is $1$~ps (FWHM).
Note that the present model leads to the identical Hall current
around $K$ and $K'$ with $\tau_z=\pm 1$.

First, we investigate the light-induced Hall effect in the weak-driving regime, 
$E_{\text{MIR}}$ $=$ 1~MV/m. 
Fig.~\ref{fig:fig1}~(a) shows the computed Hall conductivity $\sigma_{xy}$
as a function of $\mu$,
while Fig.~\ref{fig:fig1}~(b) shows the experimental 
Hall conductivity with the weakest experimental fluence $0.01$~mJ \cite{2018McIver}.
Our results confirm that the Hall conductivity 
is proportional to the laser intensity in this regime. 
Therefore, the single-photon absorption process is expected to play a dominant role
in the light-induced Hall effect.
Indeed, both the theoretical and experimental results 
consistently show a strong suppression of the Hall effect 
once $\mu$ reaches $\pm \hbar \omega_{\text{MIR}}/2$, which is when single-photon absorption becomes suppressed.
The experimental result shows a sign change of the Hall conductivity in the higher doping
regime while the theoretical result shows the same sign in the whole chemical potential range.
The sign-change feature might be understood by a negative offset due to the inverse 
Faraday effect from a substrate \cite{HERTEL2006L1}. However, since a detailed analysis of 
the substrate effect is beyond the scope of the present work, this aspect will be 
investigated in the future.

To clarify the microscopic origin of the Hall current in the weak field regime,
we perform a perturbative analysis 
which shows that the Hall current originates from population imbalance of
photocarriers in the Brillouin zone (BZ) \cite{2018satoSI}.
An excess of photocarriers is generated on one side of the Dirac cone
compared to the other, and the non-symmetric photocarrier-distribution results 
in a net Hall current. Furthermore, the perturbative analysis 
reveals that the population imbalance is induced by the interference of 
two excitation paths: one of them is the single-photon absorption process 
in the bare Dirac band (black-dashed line in Fig.~\ref{fig:fig1}~(c)),
while the other one is the single-photon absorption in the tilted Dirac band (blue-solid line in Fig.~\ref{fig:fig1}~(c)), where the tilt is induced 
by the static source drain field.
To confirm this conclusion we further compute 
the conduction band population $\rho_{cc,\boldsymbol{K}(t)}(t)=\mathrm{Tr}
\left [ |u^s_{c\boldsymbol{K}(t)}\rangle
\langle u^s_{c\boldsymbol{K}(t)}| \rho_{\boldsymbol{K}(t)}(t) \right]$ 
using instantaneous eigenstates of the Hamiltonian;
$\hat H_{\boldsymbol{K}(t)}|u^s_{b\boldsymbol{K}(t)}\rangle=
\epsilon_{b\boldsymbol{K}(t)}|u^s_{b\boldsymbol{K}(t)}\rangle$,
where $b$ denotes the band index, $v$ and $c$ for valence and conduction bands, respectively.
Figure~\ref{fig:fig1}~(d) shows the cycle-averaged conduction-band population pump dichroism.
A population imbalance along the $x$-direction is clearly observed 
under the source-drain field along the $y$-direction.
From this, we compute the population contribution to the Hall current,
$J^{POP}_H(t)$, by multiplying populations with corresponding band velocities,
\be
J^{POP}_H(t)=-\frac{2e}{\hbar (2\pi)^2}
\sum_{b=v,c}
\int d\boldsymbol{k}
\frac{\partial \epsilon_{b\boldsymbol{K}(t)}}{\hbar \partial k_x}\rho_{bb,\boldsymbol{K}(t)}.
\label{eq:pop-jh}
\ee
The result shown in Fig.~\ref{fig:fig1}~(a) accounts for most of the total Hall conductivity.
Overall our analysis demonstrates that in the weak-field regime the Hall current mainly originates from 
a population imbalance of photocarriers in the bare Dirac band.
Importantly, the population imbalance mechanism does not rely on the specific properties of
the Dirac Hamiltonian. Therefore, the mechanism is 
general and can induce anomalous Hall currents even 
in conventional semiconductors \cite{2018satoSI}.

\begin{figure}[htbp]
  \includegraphics[width=0.9\columnwidth]{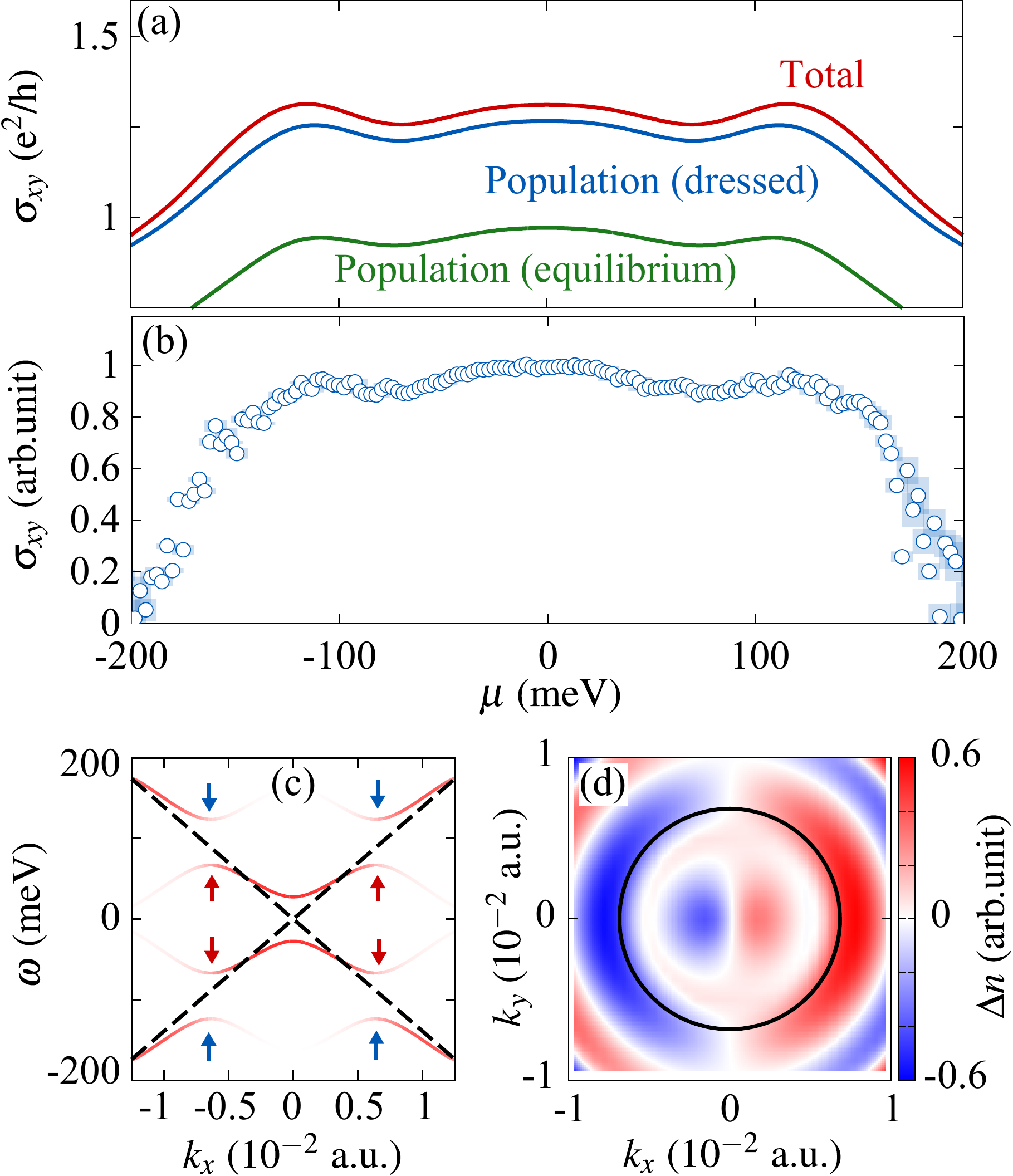}
\caption{\label{fig:fig2}
Light-induced Hall conductivity in the strong field regime,
$E_{\text{\text{MIR}}}=20$~MV/m:
(a)~The theoretical Hall conductivity $\sigma_{xy}$ 
as a function of $\mu$. The full simulation result (red),
the population contribution in the original Dirac band (green),
and the natural-orbital population contribution (blue) are shown.
(b)~The corresponding experimental Hall conductivity as in Ref.~\cite{2018McIver}.
(c)~Floquet bands (red) and
the original Dirac cone (black-dashed).
Outer (inner) edges of the resonant gap are indicated by blue (red) arrows.
(d)~Pump dichroism of natural-orbital population with source-drain bias along $y$-direction.
}
\end{figure}

Having established an asymmetric population imbalance in the Dirac bands as the source of light-induced anomalous Hall currents in the weak-field regime, we move on to explore the strong optical driving regime. We set $E_{\text{MIR}}$ to $20$~MV/m, which is the peak field intensity for the largest reported experimental fluence, $0.23~mJ$ \cite{2018McIver}. Fig.~\ref{fig:fig2}~(a) shows the computed Hall conductivity $\sigma_{xy}$ as a function of $\mu$, while Fig.~\ref{fig:fig2}~(b) shows the corresponding experimental result \cite{2018McIver}. The theoretical and experimental data are in good agreement, with both exhibiting conductivity maxima around $\mu=0$ as well as $\pm\hbar \omega_{\text{MIR}}/2$.
Fig.~\ref{fig:fig2}~(c) shows the calculated Floquet band structure for the same driving parameters. The Floquet bands show characteristic gap openings at $\epsilon_{b\boldsymbol{k}}=0$, and $\pm \hbar \omega_{\text{MIR}}/2$. Strikingly, the peaks in $\sigma_{xy}$ near $\pm \hbar \omega_{\text{MIR}}/2$ coincide with the outer edges of the Floquet bands (blue arrows in Fig.~\ref{fig:fig2}~(c)), and the width of the central plateau of $\sigma_{xy}$ is approximately the width of the gap at the Dirac point. This already indicates a close relation between the formation of Floquet-Bloch bands and the anomalous Hall effect for strong laser driving.

In order to clarify the role of Floquet-Bloch states in the generation of anomalous Hall currents, we analyze to what extent they represent the full dynamics of the system under the driving conditions. Since the complete dynamics is described by the time-dependent density matrix $\rho_{\boldsymbol{K}(t)}$, we analyze the time evolution in terms of its instantaneous eigenstates, the time-dependent natural orbitals
$|u^{NO}_{b\boldsymbol{k}}(t)\rangle$ \cite{PhysRev.97.1474},
giving
$\rho_{\boldsymbol{K}(t)}=\sum_{b=v,c}n^{NO}_{b\boldsymbol{k}}(t)
|u^{NO}_{b\boldsymbol{k}}(t)\rangle \langle u^{NO}_{b\boldsymbol{k}}(t)|$. 
This orbital basis for the dynamics gives the full dressed states of the system without the assumptions required by Floquet theory and allows us to make a detailed assessment of the Floquet picture. 

The cycle-averaged population of the natural orbitals is computed as
$\tilde n^{NO}_{b\boldsymbol{k}}=\int^{T_{\text{MIR}}}_0 dt n^{NO}_{b\boldsymbol{k}}(t)/T_{\text{MIR}}$ and
Figure~\ref{fig:fig2}~(d) shows the pump dichroism of the natural-orbital population
$\tilde n^{NO}_{b\boldsymbol{k}}$. 
Consistent with the population imbalance in the weak-field regime
(Fig.~\ref{fig:fig1}~(d)), the natural-orbital population also shows 
the imbalance along the $x$-direction. 
From the population of the natural orbitals, we can compute the dressed-band population contribution to the Hall current as
\be
J^{D-POP}_{H}=-\frac{2e}{\hbar(2\pi)^2}\sum_{b=v,c}\int d\boldsymbol{k}
\tilde v^{NO}_{b\boldsymbol{k}} \tilde n^{NO}_{b\boldsymbol{k}},
\label{eq:nat-pop-jh}
\ee
where $\tilde v^{NO}_{b\boldsymbol{k}}$ is an effective dressed-band natural-orbital velocity, 
$\tilde v^{NO}_{b \boldsymbol k}=v_F\int^{T_{\text{MIR}}}_0 dt 
\langle u^{NO}_{b \boldsymbol k}(t)|\sigma_y|u^{NO}_{b \boldsymbol k}(t)\rangle/T_{\text{MIR}}$.
The blue curve in Fig.~\ref{fig:fig2}~(a) shows the contribution
of the natural-orbital population computed by Eq.~(\ref{eq:nat-pop-jh}).
The contribution of the natural-orbital population reproduces
the full signal very well in the whole investigated chemical-potential range, indicating that a dressed state picture gives a valid description of the light-induced Hall current. 
As a reference, we also evaluated the contribution from the population imbalance 
in the bare Dirac band with Eq.~(\ref{eq:pop-jh})
and show the result as the green curve in Fig.~\ref{fig:fig2}~(a). From the results it is clear that the population imbalance in the bare Dirac band is not sufficient to describe the full signal. Thus we confirmed that the dressed states describe a significantly different electronic structure and that the light dressing enhances the light-induced Hall effect.

Having established the role of dressed bands in the strong-field regime, we turn to the relation between Floquet states and 
natural orbitals under continuous circular laser fields
without the source-drain field.
For this purpose, we introduce the \textit{Floquet fidelity}, $S_{\boldsymbol k}$,
as a measure of similarity between Floquet states $|u^{F}_{b\boldsymbol k}(t)\rangle$ 
and natural orbitals $|u^{NO}_{b\boldsymbol k}(t)\rangle$. The  Floquet fidelity is defined as the absolute value of 
the determinant of the fidelity matrix,
$S_{\boldsymbol k}=|\mathrm{det} F_{\boldsymbol k}|$,
where each element of the matrix $F_{\boldsymbol k}$ 
is the cycle-averaged overlap between natural orbitals and Floquet states,
$F_{\boldsymbol k,ij}= 
\int^{T_{\text{MIR}}}_0dt \left |\langle u^{NO}_{ik}(t)|u^{F}_{jk}(t)\rangle\right |^2 /T_{\text{MIR}}$
\cite{2018satoSI}.
The Floquet fidelity $S_{\boldsymbol k}$ can take a maximum value of one only if 
the natural orbitals and the Floquet states are identical.

\begin{figure}[htbp]
  \includegraphics[width=\columnwidth]{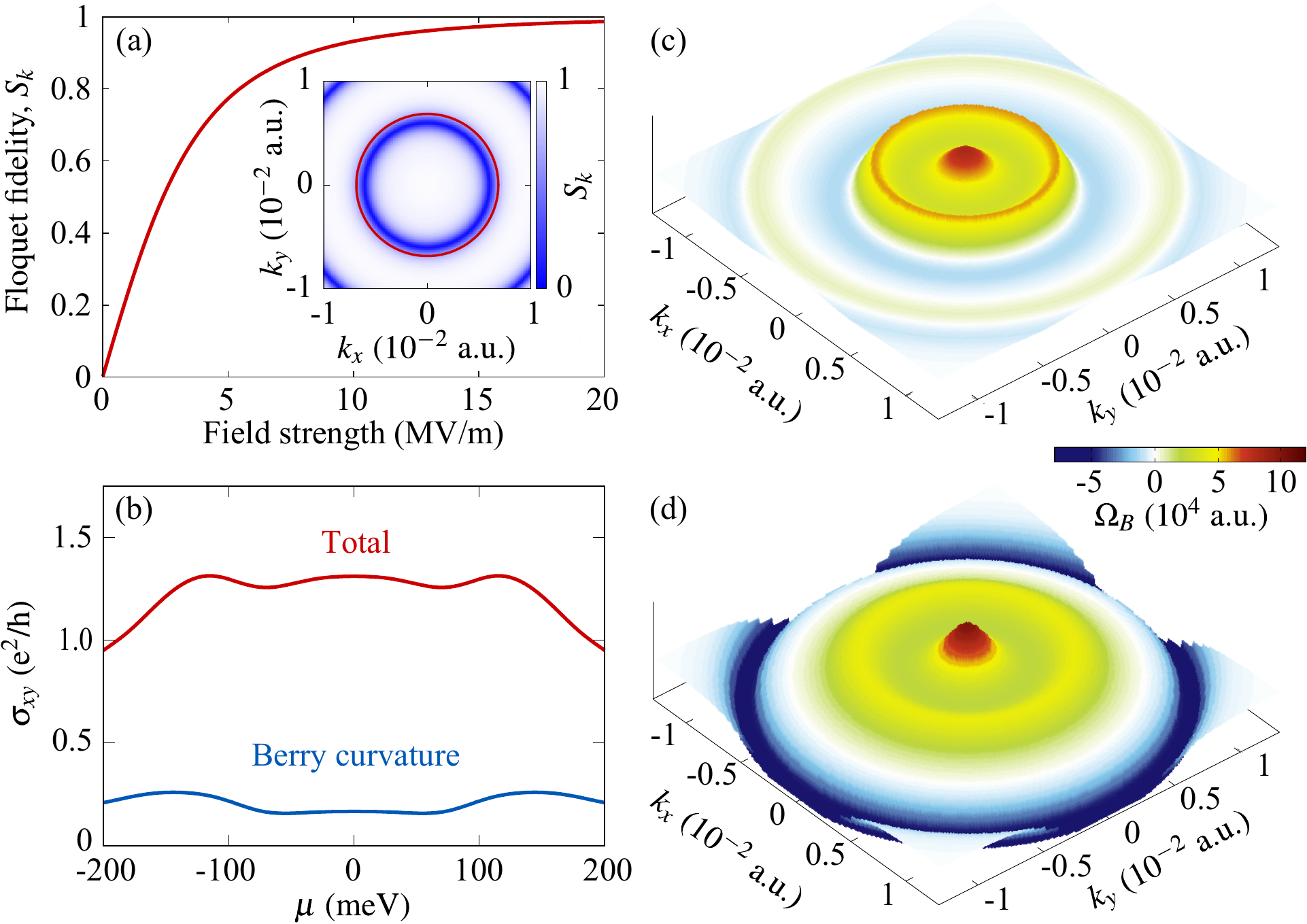}
\caption{\label{fig:fig3}
Relation of the steady state orbitals and the Floquet states:
(a)~Floquet fidelity, $S_{\boldsymbol k}$, at the Dirac point, $\boldsymbol k=0$,
as a function of the driving field strength. The inset shows
the Floquet fidelity in the BZ in the strong field regime where
$E_{\text{MIR}}=20$~MV/m.
(b) Comparison between the full conductivity $\sigma_{xy}$ and the Berry curvature
contribution $\sigma^T_{xy}$.
(c)~The Berry curvature of the steady-state natural orbitals
in the strong field regime. 
(d)~The Berry curvature of the corresponding Floquet states.
}
\end{figure}

Fig.~\ref{fig:fig3}~(a) shows the Floquet fidelity $S_{\boldsymbol k}$
at the Dirac point as
a function of the driving field strength. The system is far from the Floquet limit
in the weak-field regime ($S_{\boldsymbol k}\approx 0$), while it approaches the Floquet limit in the strong-field regime
($S_{\boldsymbol k}\approx 1$). 
This behavior is consistent with the above findings: the population imbalance
in the bare Dirac band dominates the Hall current in the weak-field regime,
while the dressed-state picture is more appropriate in the strong-field regime.

A map of the Floquet fidelity $S_{\boldsymbol k}$ in the BZ 
in the strong field regime is shown in the inset of Fig.~\ref{fig:fig3}~(a), where 
the single-photon resonance,
$\hbar v_F|\boldsymbol k|=\hbar \omega_{\text{MIR}}/2$, is indicated as a red circle.
This result shows that Floquet states are established throughout a large portion of the BZ including the Dirac point, except for a ring close to the
the single-photon resonance
($S_{\boldsymbol k}\approx 1$), where the steady state appears to be strongly disturbed ($S_{\boldsymbol k}\approx 0$).
In relation to the Floquet-bandstructure in Fig.~\ref{fig:fig2}~(c), this indicates that the circular laser field
is able to establish the outer edges of the Floquet states (the blue
arrows in Fig.~\ref{fig:fig2}~(c)), while the inner edges (red arrows)
do not form. Moreover, the realization of the outer Floquet edges 
is supported by the appearance of the peaks in Fig.~\ref{fig:fig2}~(a) that are positioned at the resonance energies.

Having demonstrated the relevance of a Floquet dressed-state picture under strong-field driving,
we now discuss the role of Berry curvature of the dressed states for the Hall current. We compute the cycle-averaged Berry curvature of the natural orbitals, 
$\Omega^{NO}_B(\boldsymbol k)=-i  \int^{T_{\text{MIR}}}_0 dt \left [\nabla_{\boldsymbol k}\times  
\langle u^{NO}_{b\boldsymbol k}(t)|\nabla_{\boldsymbol k}|u^{NO}_{b\boldsymbol k}(t)\rangle
\right]_z /T_{\text{MIR}}$ shown in
Fig.~\ref{fig:fig3}~(c), while
Fig.~\ref{fig:fig3}~(d) shows the Berry curvature of the corresponding Floquet state at $E_{\text{\text{MIR}}}=20$~MV/m.
Both the natural orbitals and the Floquet state consistently show 
a positive Berry curvature at the Dirac point ($\boldsymbol k=0$) and at the first resonance 
($v_F|\boldsymbol k|=\hbar \omega_{\text{MIR}}/2$).
Indeed the natural-orbital
Berry curvature integrated over the BZ is found to be 
$\int d\boldsymbol k\Omega^{NO}_B(\boldsymbol k)=\pm \pi$ for each Dirac cone, 
which is consistent with the topological Floquet-Chern insulator picture \cite{PhysRevB.79.081406}.

Finally we quantify the Hall currents expected solely from the time-averaged Berry curvature of the bands by computing the Hall conductivity \cite{PhysRevB.79.081406}
\be
\sigma^T_{xy}=\frac{2e^2}{\hbar}\int \frac{d\boldsymbol k}{(2\pi)^2} 
\sum_{b}n^{NO}_{b\boldsymbol k}\Omega^{NO}_{b\boldsymbol k}.
\label{eq:h-cond-topo}
\ee
The result shown in Fig.~\ref{fig:fig3}~(b) clearly demonstrates
that light-induced Berry-curvature has a non-zero contribution to the Hall conductivity.
However, this contribution is much smaller.
Furthermore, by computing the dependence of this effect on the field 
strength, we find that it changes sign~\cite{2018satoSI}.

Our modelling suggests that the recently observed light-induced anomalous Hall effect
in graphene results mainly from
the asymmetric distribution of photocarriers in the topologically non-trivial Floquet-Bloch states while their Berry curvature contribution is clearly smaller.
Nevertheless, it is striking that
the theoretical Hall conductivity saturates on the order of $\sim 2e^2/h$,
which is consistent with the experimental observation \cite{2018McIver},
despite the strong population imbalance contribution~\cite{2018satoSI}. 
This may just be a coincidence, but the possibility of hidden connections to topological invariants should be investigated further. The anomalous velocity contribution from the Berry curvature of the Floquet-Bloch bands could become larger at higher laser frequencies or for more off-resonant driving, which might be realized in other material platforms \cite{Claassen2016,Shin2018}. The population imbalance effect can also be expected to play a role in quantum simulation experiments studying Floquet effects, such as of ultracold fermions in driven optical lattices \cite{Jotzu2014,Flaschner1091,Aidelsburger2014}, as well as in other driven quantum materials. 

More broadly, these results highlight the importance of adopting a dressed-state picture in understanding the non-equilibrium electrical transport properties of optically-driven solids. They also demonstrate the viability of Floquet-engineering under experimentally realistic conditions, which opens the doors to exploring a host of other exciting non-equilibrium quantum transport phenomena. For example, Floquet-engineering of effective couplings in solids, such as dynamical Hubbard U, has recently been proposed as a means to significantly modify electronic states in correlated insulators \cite{PhysRevLett.121.097402,PhysRevLett.103.133002,PhysRevLett.115.187401} which can also induce nontrivial topology~\cite{Topp2018}. Future work could for instance explore the possibilities for light-controlled topological edge states \cite{PhysRevB.90.115423,claassen2018universal} that are opened up by the present ultrafast-transport setup and complementary ultrafast real-space imaging techniques~\cite{Cocker2016,PhysRevB.91.245155}.


This work was supported by the European Research Council (ERC-2015-AdG694097)
and the Deutsche Forschungsgemeinschaft through the SFB 925.
The Flatiron Institute is a division of the Simons Foundation.
S.A.S. gratefully acknowledges the fellowship from the Alexander von Humboldt Foundation. M.A.S. acknowledges financial support by the DFG through the Emmy Noether programme (SE 2558/2-1). P.T. acknowledges the received funding from the European Unions Horizon 2020 research and innovation programme under the Marie Sklodowska-Curie grant agreement No 793609.
M.N. acknowledges support from Stiftung der Deutschen Wirtschaft.
L.M., A.R. and A.C. acknowledge support from the Cluster of Excellence 'Advanced Imaging of Matter' (AIM)


\bibliography{ref}

\end{document}


\author{S.~A.~Sato}
\email{shunsuke.sato@mpsd.mpg.de}
\affiliation 
{Max Planck Institute for the Structure and Dynamics of Matter, Luruper Chaussee 149, 22761 Hamburg, Germany}

\author{J.~W.~McIver}
\affiliation 
{Max Planck Institute for the Structure and Dynamics of Matter, Luruper Chaussee 149, 22761 Hamburg, Germany}

\author{M.~Nuske}
\affiliation 
{Zentrum f\"ur Optische Quantentechnologien and Institut f\"ur Laserphysik,
Universit\"at Hamburg, Luruper Chaussee 149, 22761 Hamburg, Germany}

\author{P.~Tang}
\affiliation 
{Max Planck Institute for the Structure and Dynamics of Matter, Luruper Chaussee 149, 22761 Hamburg, Germany}

\author{G.~Jotzu}
\affiliation 
{Max Planck Institute for the Structure and Dynamics of Matter, Luruper Chaussee 149, 22761 Hamburg, Germany}

\author{B.~Schulte}
\affiliation 
{Max Planck Institute for the Structure and Dynamics of Matter, Luruper Chaussee 149, 22761 Hamburg, Germany}

\author{H.~H\"ubener}
\affiliation 
{Max Planck Institute for the Structure and Dynamics of Matter, Luruper Chaussee 149, 22761 Hamburg, Germany}

\author{U.~De~Giovannini}
\affiliation 
{Max Planck Institute for the Structure and Dynamics of Matter, Luruper Chaussee 149, 22761 Hamburg, Germany}

\author{L.~Mathey}
\affiliation 
{Zentrum f\"ur Optische Quantentechnologien and Institut f\"ur Laserphysik,
Universit\"at Hamburg, Luruper Chaussee 149, 22761 Hamburg, Germany}
\affiliation
{The Hamburg Centre for Ultrafast Imaging, University of Hamburg,
Luruper Chaussee 149, 22761 Hamburg, Germany}

\author{M.~A.~Sentef}
\affiliation 
{Max Planck Institute for the Structure and Dynamics of Matter, Luruper Chaussee 149, 22761 Hamburg, Germany}

\author{A.~Cavalleri}
\affiliation 
{Max Planck Institute for the Structure and Dynamics of Matter, Luruper Chaussee 149, 22761 Hamburg, Germany}

\author{A.~Rubio}
\affiliation 
{Max Planck Institute for the Structure and Dynamics of Matter, Luruper Chaussee 149, 22761 Hamburg, Germany}
\affiliation 
{Center for Computational Quantum Physics (CCQ), The Flatiron Institute, 162 Fifth Avenue, New York, NY
10010, USA}

\title{Microscopic theory for the light-induced anomalous Hall effect in graphene: Supporting Information}

\maketitle
\section{Theoretical modeling}
Here, we describe the details of our theoretical modeling of
electron dynamics in graphene under laser fields.
The electron dynamics is described by the following quantum Liouville equation
for the one-particle reduced density matrix,
\be
\frac{d}{dt}\rho_{\boldsymbol K(t)}(t) = 
\frac{\left [H_{\boldsymbol K(t)},\rho_{\boldsymbol K(t)} \right ]}{i\hbar}
+\hat D_{\boldsymbol K(t)}\rho_{\boldsymbol K(t)}(t)
\label{eq:liouville}
\ee
with the Dirac Hamiltonian
$H_{\boldsymbol K(t)}=\hbar v_F \tau_z \sigma_x K_x(t)+\hbar v_F \sigma_y K_y(t)$
and an effective relaxation operator $\hat D_{\boldsymbol K(t)}$.
Here, $v_F$ is the Fermi velocity, $\sigma_{x/y}$ is the Pauli matrix. The different
chirality of the Dirac fermions at $K$ and $K'$ points is given by $\tau_z=\pm 1$.
In this work, we employ a simple relaxation time approximation \cite{PhysRevLett.73.902}
for the relaxation operator $\hat D_{\boldsymbol K(t)}$.
To construct the relaxation operator, we first represent
the density matrix with instantaneous eigenstates of
the Hamiltonian $H_{\boldsymbol K(t)}$ as
\be
\rho_{\boldsymbol K(t)}(t) :=
\left(
    \begin{array}{cc}
      \langle u^s_{v \boldsymbol K(t)}|\hat \rho_{\boldsymbol K(t)}(t)|u^s_{v \boldsymbol K(t)}\rangle &
      \langle u^s_{v \boldsymbol K(t)}|\hat \rho_{\boldsymbol K(t)}(t)|u^s_{c \boldsymbol K(t)}\rangle  \\
      \langle u^s_{c \boldsymbol K(t)}|\hat \rho_{\boldsymbol K(t)}(t)|u^s_{v \boldsymbol K(t)}\rangle & 
      \langle u^s_{c \boldsymbol K(t)}|\hat \rho_{\boldsymbol K(t)}(t)|u^s_{c \boldsymbol K(t)}\rangle
    \end{array}
  \right) 
=
\left(
    \begin{array}{cc}
      \rho_{vv,\boldsymbol K(t)}(t) & \rho_{vc,\boldsymbol K(t)}(t) \\
      \rho_{cv,\boldsymbol K(t)}(t) & \rho_{cc,\boldsymbol K(t)}(t)
    \end{array}
  \right),
\ee
where $|u^s_{b\boldsymbol K(t)}\rangle$ is an instantaneous eigenstate
of the Hamiltoinan, 
$H_{\boldsymbol K(t)}|u^s_{b\boldsymbol K(t)}\rangle = \epsilon_{b \boldsymbol K(t)}|u^s_{b\boldsymbol K(t)}\rangle$.
Then, we construct the relaxation operator with the phenomenological
relaxation time, $T_1$ and $T_2$, in the instantaneous eigenbasis expression as follows
\be
\hat D_{\boldsymbol K(t)}\rho_{\boldsymbol K(t)} :=-
\left(
    \begin{array}{cc}
      \frac{\rho_{vv,\boldsymbol K(t)}(t)-\rho^{eq}_{v\boldsymbol K(t),\mu, T_e}}{T_1} & 
      \frac{\rho_{vc,\boldsymbol K(t)}(t)}{T_2} \\
      \frac{\rho_{cv,\boldsymbol K(t)}(t)}{T_2} &
      \frac{\rho_{cc,\boldsymbol K(t)}(t)-\rho^{eq}_{c\boldsymbol K(t),\mu, T_e}}{T_1}
    \end{array}
  \right), 
\ee
where $\rho^{eq}_{b,\mu,T_e}$ is the Fermi-Dirac distribution,
\be
\rho^{eq}_{b,\mu,T_e} = \frac{1}{e^{\left (\epsilon_{b\boldsymbol K(t)}-\mu \right)/k_B T_e}+1}.
\ee

In this work, the decoherence time $T_2$ is set to $20$~fs according to
the electron-electron scattering time scale, while 
and the population relaxation time $T_1$ is set to $100$~fs 
according the electron thermalization time scale 
\cite{PhysRevB.83.153410,Brida2013,PhysRevLett.115.086803}.
However, as will be shown in the following section,
the qualitative behaviors of the light-induced Hall effect do not much
depend on the choice of the relaxation time.

\section{Evaluation of the light-induced Hall conductivity}

In order to investigate the light-induced Hall effect,
we compute the electron dynamics under a circular mid-infrared (MIR)
laser pulse $\boldsymbol E_{MIR}(t)$
and a static source-drain field $\boldsymbol E_{SD}(t)$.
In this work, the pulse width of the MIR field is set to $1$~ps,
and the wavelength is set to $6.5$~$\mu$m.
The corresponding mean photon energy is about $\hbar \omega_{MIR}\approx190$~meV.
Furthermore, we assume that the MIR laser fields propagate along the 
$z$-axis, and the field polarizations are always on the $x-y$ plane.
For the source-drain field, to prevent an artificial excitation,
we employ the following smooth switching
\be
\boldsymbol E_{SD}(t) = 
\left\{
  \begin{array}{@{}ll@{}}
    E_{SD} \boldsymbol e_y , & T_{switch}<t \\
    E_{SD} \boldsymbol e_y\left [
      3\left( \frac{t}{T_{switch}}\right)^2
        -2\left( \frac{t}{T_{switch}}\right)^3
      \right ]
 , & 0<t\le T_{swtich} \\
    0 , & \text{otherwise} 
  \end{array}\right.
\ee
In this work, we set the source-drain direction to the $y$-direction,
the source-drain field strength $E_{SD}$ to 
$10^4$~MV/m, and the switching time $T_{switch}$ to $20$~fs. 

To evaluate the Hall current, we compute the current along the $x$-direction
under the presence of the circular MIR laser field and
the source-drain field. 
Following the same analysis of the experiment \cite{2018McIver},
we compute two kinds of current, $J^{(\circlearrowright)}_x(t)$
and $J^{(\circlearrowleft)}_x(t)$:
$J^{(\circlearrowright)}_x(t)$ is induced by the right-handed circular 
laser $E^{(\circlearrowright)}_{MIR}(t)$,
while $J^{(\circlearrowleft)}_x(t)$ is induced by the left-handed circular 
laser $E^{(\circlearrowleft)}_{MIR}(t)$.
Then, we define the difference of $J^{(\circlearrowright)}_x(t)$
and $J^{(\circlearrowleft)}_x(t)$ by
$\Delta J_x(t)= \left [
J^{(\circlearrowright)}_x(t)-J^{(\circlearrowleft)}_x(t)
\right ]/2
$.
So far, the current $\Delta J_x(t)$ contains high-frequency components, which
are not relevant for the transport property since the time-average of 
the high frequency component becomes zero and there is no net charge transfer.
To cleanly extract the transport property, we remove the irrelevant
high-frequency component by the temporal average and define
the theoretical Hall current as
\be
J_H(t) = \frac{1}{\sqrt{2\pi\sigma^2}}\int^{\infty}_{-\infty}dt'
e^{-\frac{(t'-t)^2}{2\sigma^2}}\Delta J_x(t'),
\ee
where the width of the window $\sigma$ is set to $100$~fs, which is
substantially longer than the optical cycle but shorter than
the pulse width.
Furthermore, we define the Hall conductivity $\sigma_{xy}$ as the ratio
of the peak Hall current $J_H(t_{peak})$ and the source-drain field strength,
\be
\sigma_{xy} = \frac{J_H(t_{peak})}{E_{SD}}.
\ee

\section{Relaxation time dependence}

\begin{figure}[htbp]
\centering
\includegraphics[width=0.6\columnwidth]{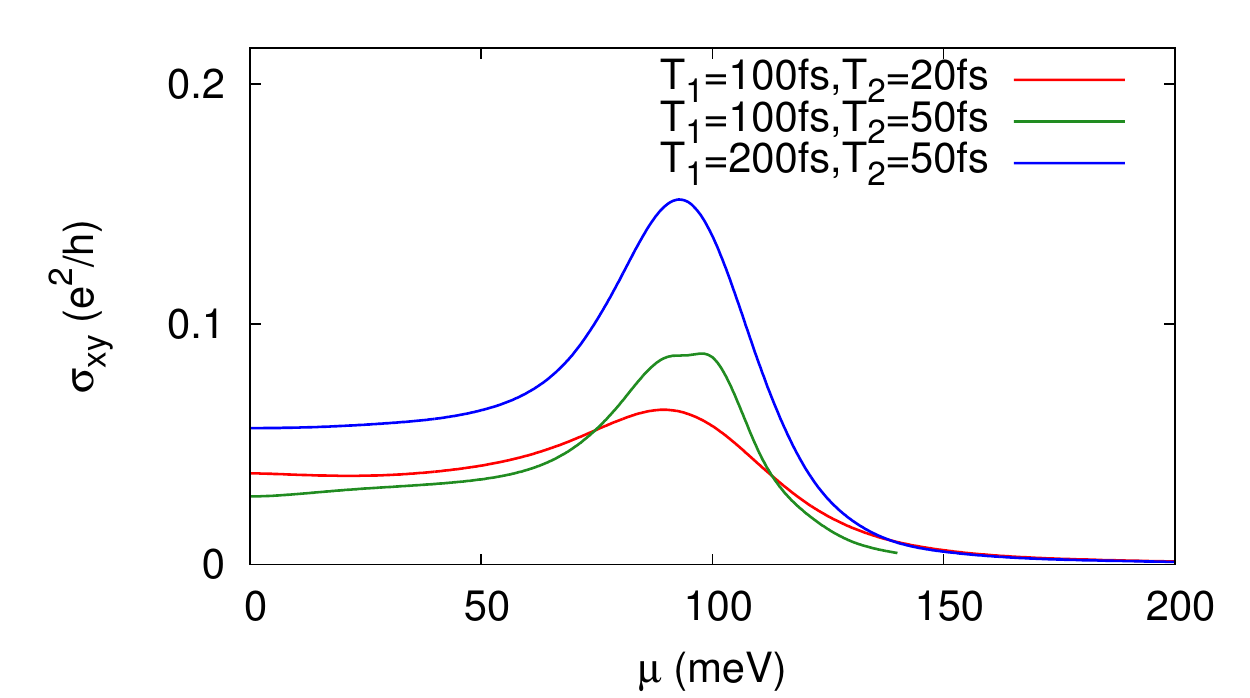}
\caption{\label{fig:relaxation_weak}
  The Hall conductivity $\sigma_{xy}$ as a function of chemical potential
  $\mu$ in the weak field regime.
  The results with different relaxation time, $T_1$ and $T_2$, are shown.
}
\end{figure}

\begin{figure}[htbp]
\centering
\includegraphics[width=0.6\columnwidth]{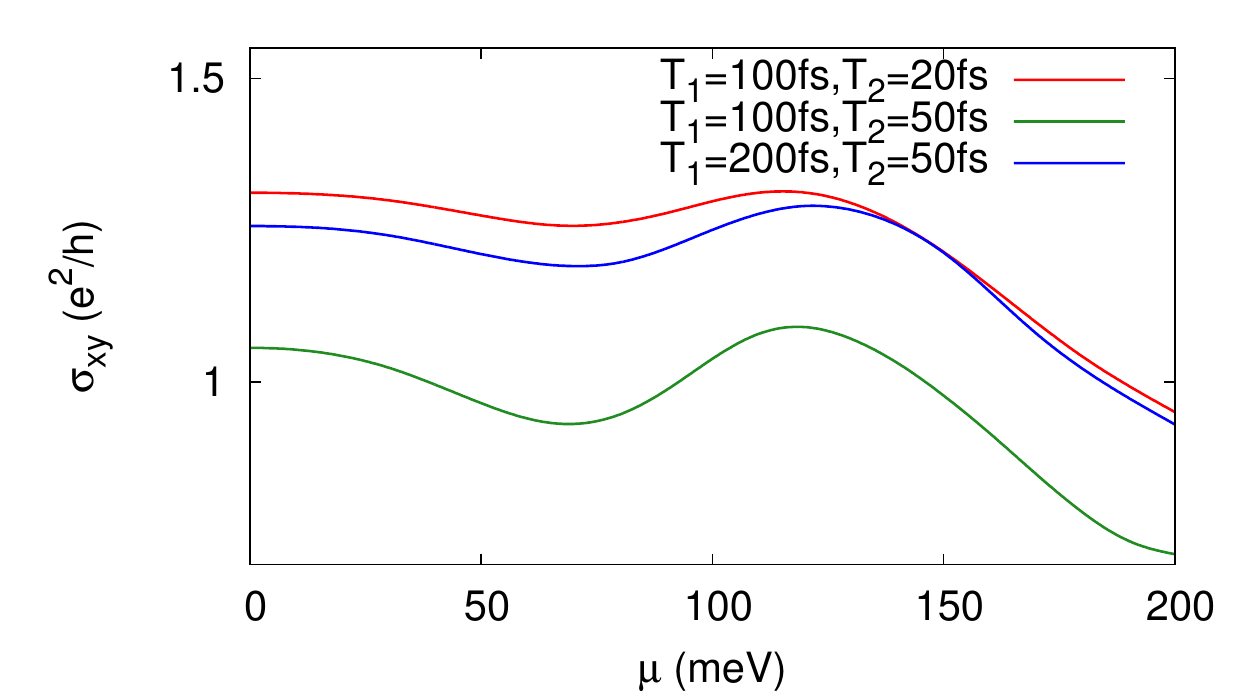}
\caption{\label{fig:relaxation_strong}
  The Hall conductivity $\sigma_{xy}$ as a function of chemical potential
  $\mu$ in the strong field regime.
  The results with different relaxation time, $T_1$ and $T_2$, are shown.  
}
\end{figure}

Here, we explore the effect of the relaxation time, $T_1$ and $T_2$
in the light-induced Hall effect.
Based on the above procedure, we compute
the Hall conductivity $\sigma_{xy}$ as a function of chemical potential $\mu$.
Figure~\ref{fig:relaxation_weak} shows the computed Hall
conductivity $\sigma_{xy}$ with the different relaxation time, $T_1$ and
$T_2$, in the weak field regime, where the field strength
of the circular laser is set to $E_{MIR}=1$~MV/m.
The same results in the strong field regime ($E_{MIR}=20$~MV/m)
are shown in Fig.~\ref{fig:relaxation_strong}.
One can clearly confirm that the qualitative behaviors
of the Hall conductivity do not depend on the choice of
the relaxation time in both the weak and the strong field regimes.

\section{Comparison of pulsed and continuous-wave laser fields}

Because the relaxation times, $T_1$ and $T_2$, are much shorter
than the pulse width in this work,
the system is expected to realize a steady state
due to the balance between the laser-excitation and the relaxation.
To confirm this fact, we compute the electron dynamics
under a continuous-wave circular laser field instead of a laser pulse,
and evaluate the Hall conductivity after the system reaches the steady state.

\begin{figure}[htbp]
\centering
\includegraphics[width=0.6\columnwidth]{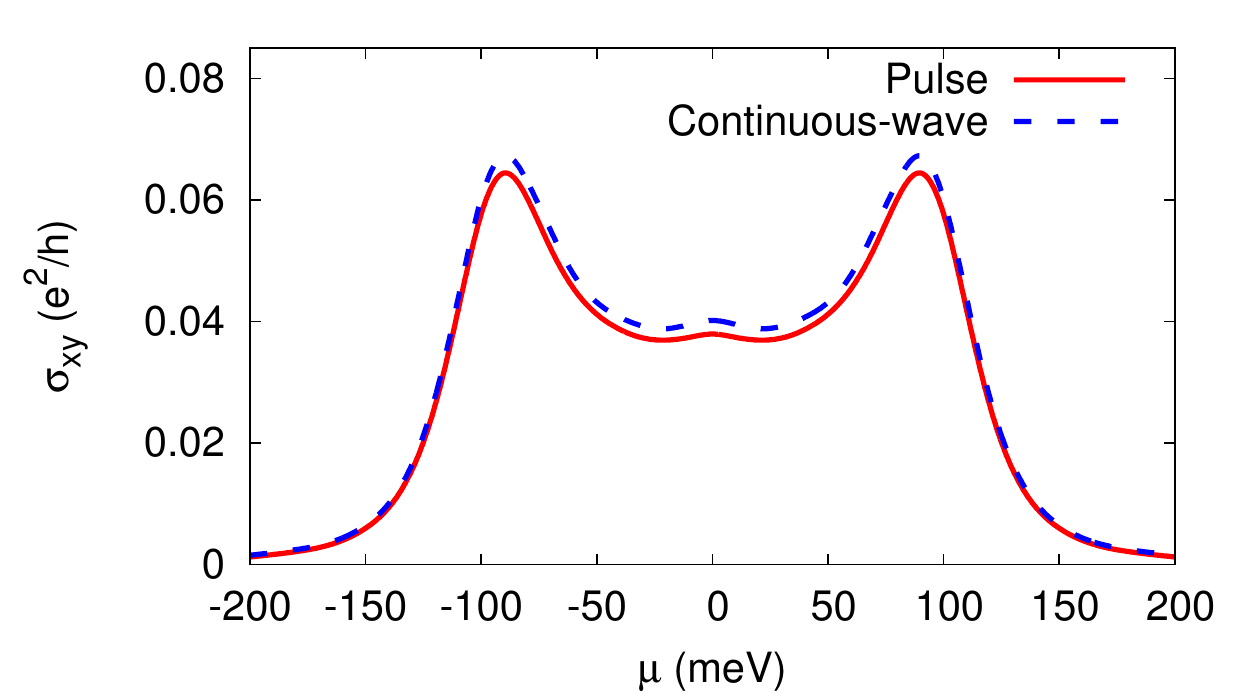}
\caption{\label{fig:cw_weak}
  The Hall conductivity $\sigma_{xy}$ as a function of chemical potential
  $\mu$ in the weak field regime.
  The results computed with the laser pulse (red-solid) and
  the continuous-wave laser (blue-dashed) are shown.
}
\end{figure}

\begin{figure}[htbp]
\centering
\includegraphics[width=0.6\columnwidth]{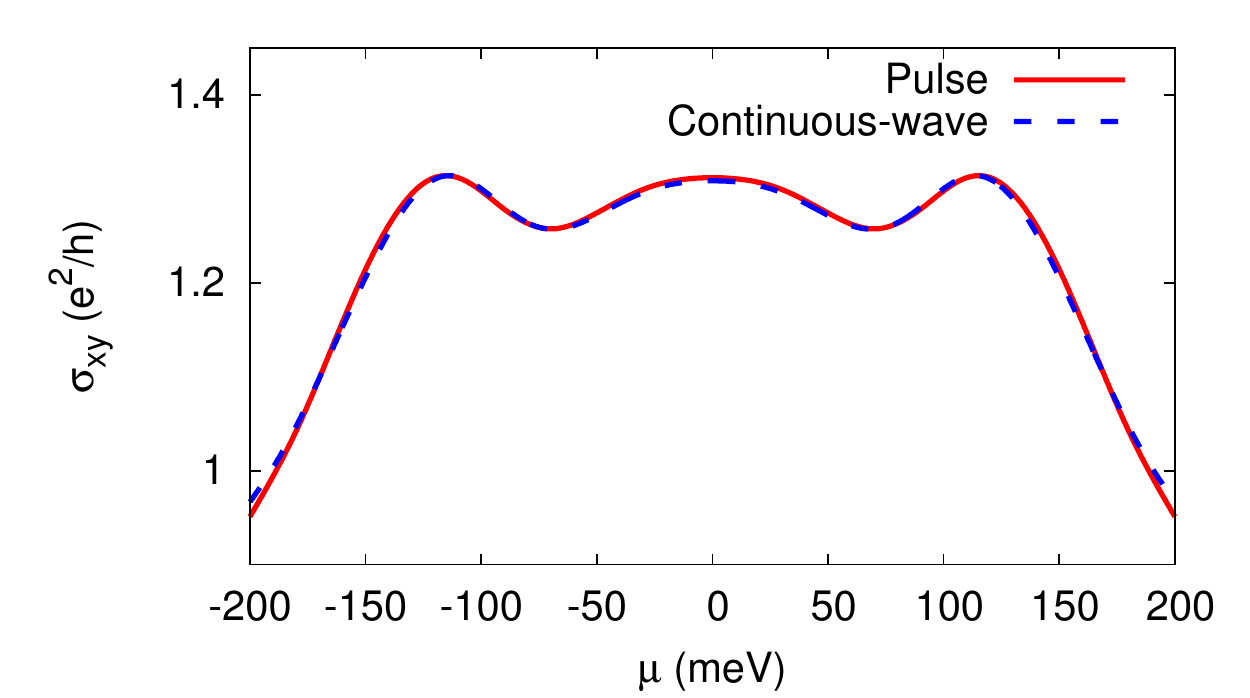}
\caption{\label{fig:cw_strong}
  The Hall conductivity $\sigma_{xy}$ as a function of chemical potential
  $\mu$ in the weak field regime.
  The results computed with the laser pulse (red-solid) and
  the continuous-wave laser (blue-dashed) are shown.
}
\end{figure}

Figure~\ref{fig:cw_weak} shows the light-induced Hall conductivities $\sigma_{xy}$ 
evaluated with a laser pulse and a continuous-wave in the weak field regime,
where the peak field strength is set to $1$~MV/m.
The same comparison in the strong field regime ($E_{MIR}=20$~MV/m)
is shown in Fig.~\ref{fig:cw_strong}.
As seen from both figures, the Hall conductivity
evaluated with a laser pulse is almost perfectly reproduced 
by that evaluated with the continuous-wave laser field.
Therefore, we clearly confirmed that
the Hall conductivity evaluated by the laser pulse reflects
the property of the steady state that is realized
by the balance of the laser-excitation and the relaxation.

\section{Perturbation analysis of the light-induced Hall effect in the Dirac band
\label{sec:pert-dirac}}
To provide microscopic insight into the light-induced Hall effect in graphene,
we apply a perturbative analysis.
For simplicity, we ignore the relaxation effect, and thus, the system is described
by the following Schr\"odinger equation instead of Eq.~(\ref{eq:liouville}),
\be
i\hbar \frac{d}{dt}|\psi_{\boldsymbol k}(t)\rangle
=H_{\boldsymbol K(t)}|\psi_{\boldsymbol k}(t)\rangle.
\label{eq:tdse}
\ee

We perform the perturbative analysis based on
the Houston state expansion \cite{PhysRev.57.184,PhysRevB.33.5494},
\be
|\psi_{\boldsymbol k}(t)\rangle
&=&c_{v\boldsymbol k}(t)e^{i\gamma_{v\boldsymbol k}(t)}
\mathrm{exp}\left[
-\frac{i}{\hbar}\int^t_0dt'\epsilon_{v\boldsymbol K(t')}
\right]|u^s_{v\boldsymbol K(t)}\rangle \nonumber \\
&+&c_{c\boldsymbol k}(t)e^{i\gamma_{c\boldsymbol k}(t)}
\mathrm{exp}\left[
-\frac{i}{\hbar}\int^t_0dt'\epsilon_{c\boldsymbol K(t')}
\right]|u^s_{c\boldsymbol K(t)}\rangle,
\ee
where $|u~s_{b\boldsymbol K(t)}\rangle$ is an eigenstate
of the instantaneous Hamiltonian, $H_{\boldsymbol K(t)}$,
$\epsilon_{b \boldsymbol K(t)}$ is its eigenvalue,
$c_{b\boldsymbol k}(t)$ is an expansion coefficient,
and $\gamma_{b\boldsymbol k}(t)$ is a geometrical phase defined by
\be
\gamma_{b\boldsymbol k}(t)=-i\int^t_0dt'
\langle u^s_{b\boldsymbol K(t)}|\frac{d}{dt'}| u^s_{b\boldsymbol K(t)}\rangle.
\ee

For practical calculation, we assume the following forms for 
the the instantaneous eigenstates,
\be
|u^s_{v,\boldsymbol K(t)}\rangle = \frac{1}{\sqrt{2}}
\left(
    \begin{array}{c}
      1  \\
      -\frac{\tau_z K_x(t)+iK_y(t)}
      {\sqrt{K^2_x(t)+K^2_y(t)}}
    \end{array}
  \right), 
\ee
and 
\be
|u^s_{c,\boldsymbol K(t)}\rangle = \frac{1}{\sqrt{2}}
\left(
    \begin{array}{c}
      1  \\
      +\frac{\tau_z K_x(t)+iK_y(t)}
      {\sqrt{K^2_x(t)+K^2_y(t)}}
    \end{array}
  \right).
\ee

Then, one can obtain the Schr\"odinger equation for the coefficient vectors,
\be
i\hbar \frac{d}{dt} \boldsymbol{c}_{\boldsymbol k}(t)=\frac{e}{2}
\frac{\tau_z}{\boldsymbol K^2(t)}\left [
\boldsymbol K(t) \times \boldsymbol E(t)
\right ]_z
\left(
    \begin{array}{cc}
      0 & e^{-\frac{2\hbar v_F}{i\hbar}\int^t_0dt'|\boldsymbol K(t')|} \\
      e^{\frac{2\hbar v_F}{i\hbar}\int^t_0dt'|\boldsymbol K(t')|} & 0
    \end{array}
  \right)\boldsymbol{c}_{\boldsymbol k}(t),
  \label{eq:tdse-houston-basis-dirac}
\ee
where the coefficient vector $\boldsymbol{c}_{\boldsymbol k}(t)$ is defined 
as
\be
\boldsymbol{c}_{\boldsymbol k}(t) =
\left(
    \begin{array}{cc}
      c_{c,\boldsymbol k}(t)   \\
      c_{v,\boldsymbol k}(t)
    \end{array}
  \right).
\ee

Then, we elucidate the nonlinear carrier-injection by the circular laser field 
and the source-drain field.
For this purpose, we assume the following form for the electric field,
\be
\boldsymbol E(t) = \boldsymbol e_x E^{P}_x(t) + \boldsymbol e_y E^{P}_x(t) + \boldsymbol e_x E^{DC},
\ee
where $E^{P}_{x/y}$ denotes the pump electric field for $x/y$-direction, and
$E^{DC}$ denotes the static electric field for $x$-direction.

Thanks to the structure of the Hamiltonian in Eq.~(\ref{eq:tdse-houston-basis-dirac}),
the coefficient vector, $\boldsymbol{c}_{\boldsymbol k}(t)$, can be
accurately evaluated up to the second-order of the electric fields as
\be
\boldsymbol{c}_{\boldsymbol k}(t)=
\frac{1}{i\hbar}
\int^t_0 dt'
\frac{e}{2}
\frac{\tau_z}{\boldsymbol K^2(t')}
\left [
\boldsymbol K(t') \times \boldsymbol E(t')
\right ]_z
\left(
    \begin{array}{cc}
      0 & e^{-\frac{2\hbar v}{i\hbar}\int^{t'}_0dt" \left| \boldsymbol K(t")\right|}  \\
      e^{\frac{2\hbar v}{i\hbar}\int^{t'}_0dt" \left| \boldsymbol K(t")\right|} & 0
    \end{array}
  \right)
\left(
    \begin{array}{cc}
      0   \\
      1
    \end{array}
  \right). \nonumber \\
\ee

The conduction component can be written as,
\be
c_{+,\boldsymbol k}(t\rightarrow \infty) = 
\frac{1}{i\hbar}
\int^{\infty}_0 dt'
\frac{e}{2}
\frac{\tau_z}{\boldsymbol K^2(t')}
\left [
\boldsymbol K(t') \times \boldsymbol E(t')
\right ]_z
e^{-\frac{2\hbar v}{i\hbar}\int^{t'}_0dt" \left| \boldsymbol K(t")\right|}.
\label{eq:coeff-c-hall-01}
\ee

For simplicity, we assume that (i) the Fourier transform
of the pump electric field $|\tilde E^P_{x/y}(\omega)|$ is localized
in the frequency domain around $\hbar \omega=\hbar|\boldsymbol k|v_F$,
and (ii) $\boldsymbol k$ is on $y$-axis; $k_x=0$.
Ignoring the third and higher order contributions in 
Eq.~(\ref{eq:coeff-c-hall-01}), we obtain
\be
c_{+,\boldsymbol k}(t\rightarrow \infty) &=& 
-\frac{\tau_z}{i\hbar}\frac{e}{2}\frac{k_y}{k^2}
\int^{\infty}_0 dt'E^P_x(t')e^{2i vkt'}
 \nonumber \\
&+&
\frac{\tau_z}{i\hbar}\frac{e}{2}
\frac{e}{\hbar}\frac{1}{k^2}
\int^{\infty}_0 dt^\prime
\frac{A^{DC}(t^\prime)}{c}E^P_y(t^\prime)
e^{2i vkt'} \nonumber \\
&+&
\frac{\tau_z}{i\hbar}
\frac{e}{2}\frac{e}{\hbar}
\frac{1}{k^2}
E^{DC}_x
\int^{\infty}_0 dt^\prime
\frac{A^P_y(t^\prime)}{c}
e^{2i vkt^\prime}\nonumber \\
&-&
\frac{\tau_z}{i\hbar}\frac{e}{2}\frac{e}{\hbar}
\frac{2iv}{k}E^{DC}_x
e^{2i vkt^\prime}
\int^{t^\prime}_0dt^{\prime\prime}\frac{A^{P}(t^{\prime\prime})}{c},
\label{eq:coeff-c-hall-01}
\ee
where $A^{P}_{x/y}$ and $A^{DC}$ are amplitude of the vector potentials
corresponding to the electric fields $E^{P}_{x/y}$ and $E^{DC}$ ,respectively.

In the right hand side of Eq.~(\ref{eq:coeff-c-hall-01}), 
the first term corresponds to the one photon absorption process
by the pump light with the first order perturbation.
The second term corresponds to the nonlinear excitation due to 
the coupling of the pump-induced inter-band transition
and the DC-field-induced intra-band acceleration.
The third and the last terms correspond to the nonlinear coupling
of the pump-induced intra-band acceleration and the DC-field-induced inter-band 
transition.
Therefore, we may classify the first term as the \textit{direct} resonant excitation,
while the rest terms as the \textit{indirect} resonant excitation
assisted by the source-drain field.

One may further proceed with the evaluation as
\be
c_{+,\boldsymbol k}(t\rightarrow \infty) &=& 
-\frac{\tau_z}{i\hbar}
\frac{e}{2}
\frac{k_y}{k^2}\tilde E^P_x(2vk_y)
\nonumber \\
&+&
\frac{\tau_z}{\hbar}\frac{e}{2}\frac{e}{\hbar}
\frac{1}{k^2}E^{DC}_x \frac{\partial}{\partial \omega}
\tilde E^P_y(\omega) \Bigg |_{\omega=2vk}
\nonumber \\
&-&
\frac{\tau_z}{\hbar}
\frac{e}{2}\frac{e}{\hbar}
\frac{1}{k^2}\frac{1}{vk}E^{DC}_x
\tilde E^P_y(2vk).
\label{eq:coeff-c-hall-02}
\ee
Here, as explained above, the first term corresponds to the resonant excitation
by the pump pulse [the first term of Eq.~(\ref{eq:coeff-c-hall-02})],
while the other terms correspond to the nonlinear photocarrier injection
assisted by the static electric field.

Assuming perfect circular pump laser,
$\tilde E^P(\omega) \equiv \tilde E^P_x(\omega) =\pm i \tilde E^P_y(\omega)$,
the injected population on $y$-axis can be evaluated up to the first order of $E^{DC}$ as
\be
n_c(k_y,k_x=0) &=& \left|c_{+,\boldsymbol k}(t\rightarrow \infty) \right|^2
\nonumber \\
&\approx&
\left (\frac{1}{\hbar}\frac{e}{2} \right )^2\frac{1}{k^2}
\left |\tilde E^P(\omega=2vk) \right|^2 \nonumber \\
&\mp&
\left (\frac{1}{\hbar}\frac{e}{2} \right )^2 \frac{e}{\hbar} \frac{k_y}{k^4}
E^{DC}_x \frac{\partial}{\partial \omega} |\tilde E^P(\omega)|^2 \Bigg|_{\omega=2vk}
\nonumber \\
&\pm& 
\left (\frac{1}{\hbar}\frac{e}{2} \right )^2 \frac{e}{\hbar} \frac{k_y}{k^4}
\frac{2}{vk}E^{DC}_x |\tilde E^P(\omega=2vk)|^2
\ee
Here, the upper sign corresponds to the right-handed circular pump,
while the lower sign corresponds to the left-handed. Therefore,
the population difference between the right- and the left-handed circular pump
becomes
\be
\Delta n_c(k_y,k_x=0) &=& 
n^{(\circlearrowright)}_c(k_y,k_x=0)
-n^{(\circlearrowleft)}_c(k_y,k_x=0) \nonumber \\
&=&-
2\left (\frac{1}{\hbar}\frac{e}{2} \right )^2 \frac{e}{\hbar} \frac{k_y}{k^4}
E^{DC}_x \frac{\partial}{\partial \omega} |\tilde E^P(\omega)|^2 \Bigg|_{\omega=2vk}
\nonumber \\
&&+
4\left (\frac{1}{\hbar}\frac{e}{2} \right )^2 \frac{e}{\hbar} \frac{k_y}{k^4}
\frac{1}{vk}E^{DC}_x |\tilde E^P(\omega=2vk)|^2.
\label{eq:population-diff-circular}
\ee

One sees that the population difference of Eq.~(\ref{eq:population-diff-circular})
breaks the symmetry for $k_y$ direction.
Therefore, the population imbalance is clearly formed along the $y$-direction
under the static field along the $x$-direction.
Note that the population imbalance is formed by the interference
between the direct resonant excitation [the first term of Eq.~(\ref{eq:coeff-c-hall-02})]
and the nonlinear photocarrier injection 
assisted by the presence of the source-drain field
[the other terms of Eq.~(\ref{eq:coeff-c-hall-02})].

Because the population distribution has a direct contribution to the intraband component
of current, the population imbalance along the $y$-direction
in the momnetum space under the static field along the $x$-direction
results in the net Hall current.

\subsection{Tilting of Dirac bands under a static field}

To understand the role of the static electric field
in the formation of the population imbalance in Eq.~(\ref{eq:population-diff-circular}),
we elucidate the modification of the electronic structure
induced purely by the static electric field.
For this purpose, we reconsider the time-dependent Schr\"odinger 
equation~(\ref{eq:tdse}) with the following ansatz,
\be
|\psi_{\boldsymbol k}(t) \rangle = d_{v,\boldsymbol k}(t)|u^s_{v\boldsymbol K(t)}\rangle
+d_{c,\boldsymbol k}(t)|u^s_{c\boldsymbol K(t)}\rangle.
\ee

The coefficient vector $\boldsymbol{d}_{\boldsymbol k}(t)$ satisfies the following
Schr\"odinger equation under a static electric field $\boldsymbol E_0$,
\be
i\hbar \frac{d}{dt}\boldsymbol{d}_{\boldsymbol k}(t)
&=&\hbar v|\boldsymbol K(t)|\left(
    \begin{array}{cc}
      1 &  0  \\
      0 & -1
    \end{array}
  \right)
  \boldsymbol{d}_{\boldsymbol k}(t) \nonumber \\
&+& \frac{\tau_z}{2}\frac{e}{K^2} \left [
\boldsymbol K \times \boldsymbol E_0
\right ]_z
\left(
    \begin{array}{cc}
      -1 &  1  \\
      1 & -1
    \end{array}
  \right)  \boldsymbol{d}_{\boldsymbol k}(t).
\ee

Then, we evaluate the eigenvalues of the above effective Hamiltonian
and obtain
\be
\tilde \epsilon_{c/v,\boldsymbol K}(t)
=\pm \sqrt{
\left (\hbar v |\boldsymbol K(t)| \right)^2
+\left ( 
\frac{\tau_z e}{2K^2} \left [
\boldsymbol K \times \boldsymbol E_0
\right ]_z
\right )^2
}-\frac{\tau_z e}{2K^2}\left [
\boldsymbol K \times \boldsymbol E_0
\right ]_z.
\ee

In the weak field limit, the eigenvalues can be approximated by
\be
\tilde \epsilon_{c/v,\boldsymbol K(t)} = \pm \hbar v_F |\boldsymbol K(t)|
-\frac{\tau_z e}{2K^2}\left [
\boldsymbol K \times \boldsymbol E_0
\right ]_z.
\ee
The first term is nothing but the energy of the bare Dirac band, and the second term
is the modification due to the applied static electric field.
One sees that the modification induces the distorted tilt to the bare Dirac band.
Therefore, we can conclude that the static electric field assists the nonlinear photocarrier
injection with the pump pulse by titling the Dirac band.

\section{Perturbation analysis of the light-induced Hall effect in a parabolic two-band semiconductor}

In this section, we investigate the light-induced Hall effect in a simple parabolic two-band
model in order to demonstrate that the population effect discussed in 
Sec.~\ref{sec:pert-dirac} is rather general mechanism.

\subsection{Parabolic two-band model \label{subsec:2band}}

First, we construct a parabolic two-band model.
For this purpose, we start from the following one-body Schr\"odinger equation,
\be
i \hbar \frac{\partial}{\partial t}u_{b\boldsymbol k}(\boldsymbol r,t)&=&
\left [
\frac{1}{2m}\left \{
\boldsymbol p + \hbar \boldsymbol k + \frac{e}{c}\boldsymbol A(t)
\right \}^2 + v(\boldsymbol r)
\right ]u_{b\boldsymbol k}(\boldsymbol r,t) \nonumber \\
&=&\hat h_{\boldsymbol K(t)}u_{b\boldsymbol k}(\boldsymbol r,t),
\label{eq:tdse-bloch}
\ee
where $u_{b\boldsymbol k}(\boldsymbol r,t)$ is a time-dependent Bloch state, and $v(\boldsymbol r)$
is a one-body potential that has the same periodicity as the crystal. Here, $b$
denotes a band index, while $\boldsymbol k$ denotes the Bloch wave number.
We note that the crystal momentum is shifted by the vector potential
as $\boldsymbol K(t)=\boldsymbol k + e\boldsymbol A(t)/\hbar c$.

Then, we introduce the Houston state \cite{PhysRev.57.184,PhysRevB.33.5494}
as a solution of the Schr\"odinger equation~(\ref{eq:tdse-bloch}) in the adiabatic limit;
\be
u^H_{b\boldsymbol k}(\boldsymbol r,t)=\exp \left [
-\frac{i}{\hbar}\int^t_0dt' \epsilon_{b\boldsymbol K(t')}
\right ]u^S_{b \boldsymbol K(t)},
\ee
where $ \epsilon_{b \boldsymbol K(t)}$ and $u^S_{b\boldsymbol K(t)}(\boldsymbol r)$ are an eigenvalue and
the eigenstate of the instantaneous Hamiltonian, $\hat h_{\boldsymbol K(t)}$, respectively;
\be
\hat h_{\boldsymbol K(t)}u^S_{b\boldsymbol K(t)}(\boldsymbol r) = 
\epsilon_{b\boldsymbol K(t)}u^S_{b\boldsymbol K(t)}(\boldsymbol r).
\ee

Here, we assume the following condition for the diagonal element of
the $k$-derivative operator for all bands $b$ and all $\boldsymbol k$;
\be
\int_{\Omega} d \boldsymbol r u^{S,*}_{b\boldsymbol k}(\boldsymbol r)
\frac{\partial}{\partial \boldsymbol k} u^{S}_{b\boldsymbol k}(\boldsymbol r) = 0.
\ee
This condition guarantees that there is no Berry curvature
at all $\boldsymbol k$.

To construct a two-band model, we assume that the wavefunction at each $k$-point
can be expanded by only two Houston states; one representing a valance, and
the other a conduction state;
\be
u_{\boldsymbol k}(\boldsymbol r,t)=c_{v\boldsymbol k}(t)u^H_{v\boldsymbol k}(\boldsymbol r,t)
+c_{c\boldsymbol k}(t)u^H_{c\boldsymbol k}(\boldsymbol r,t).
\label{eq:wf-expand-houston}
\ee

Inserting Eq. (\ref{eq:wf-expand-houston}) into Eq. (\ref{eq:tdse-bloch}),
one can derive an equation of motion for the coefficients $c_{v\boldsymbol k}(t)$ 
and $c_{c\boldsymbol k}(t)$,

\be
i\hbar
\frac{d}{dt}
\left(
    \begin{array}{c}
      c_{v\boldsymbol k}(t)   \\
      c_{c\boldsymbol k}(t)
    \end{array}
  \right)
=
\left(
    \begin{array}{cc}
      0 & h_{vc,\boldsymbol k}(t)  \\
      h^*_{vc, \boldsymbol k}(t) & 0
    \end{array}
  \right)
\left(
    \begin{array}{c}
      c_{v\boldsymbol k}(t)   \\
      c_{c\boldsymbol k}(t)
    \end{array}
  \right),
\nonumber \\
\label{eq:schrodinger-2band}
\ee
where the off-diagonal matrix element is given by
\be
h_{vc,\boldsymbol k}(t) = -\frac{i \boldsymbol p_{vc,\boldsymbol K(t)} \cdot \boldsymbol E(t) }
{\epsilon_{v, \boldsymbol K(t)} -\epsilon_{c, \boldsymbol K(t)}}\frac{e \hbar}{m}
e^{\frac{1}{i\hbar}\int^t dt' \left \{
\epsilon_{c, \boldsymbol K(t')} - \epsilon_{v, \boldsymbol K(t')}
\right \}
},
\nonumber \\
\ee
and
\be
\boldsymbol p_{vc,\boldsymbol K(t)} = \int_{\Omega} d\boldsymbol r
u^{S,*}_{v\boldsymbol K(t)}(\boldsymbol r) \boldsymbol p u^{S}_{c\boldsymbol K(t)}(\boldsymbol r),
\ee
where $\Omega$ is the volume of the unit-cell.
Note that Eq.~(\ref{eq:schrodinger-2band}) is nothing but 
the Houston state expansion of the Schr\"odinger equation
\cite{PhysRev.57.184,PhysRevB.33.5494} with only two Houston states.

To further simplify the model, we approximate the electronic structure
by the parabolic bands as
\be
\epsilon_{v,\boldsymbol k} &=& - \frac{\hbar^2 \boldsymbol k^2}{2m_{v}}, \\
\epsilon_{c,\boldsymbol k} &=& \epsilon_g + \frac{\hbar^2 \boldsymbol k^2}{2m_{c}},
\ee
where $\epsilon_g$ is the band gap, and $m_v$ and $m_c$ are the effective masses for
valence and conduction bands, respectively.
Here, we also define the reduced mass $\mu^{-1}=m^{-1}_v+m^{-1}_c$.

\subsection{Perturbation analysis for light-induced Hall current \label{subsec:pert-2band}}

Here, we investigate the light-induced Hall current in the parabolic two-band model
with the perturbation theory.
We set the initial wavefunction to the valence state. 
Thanks to the structure of the Schr\"odinger equation~(\ref{eq:schrodinger-2band}),
the time-dependent conduction coefficient, $c_{c\boldsymbol k}(t)$, can be 
accurately described by the following expression upto the second order of 
the electric field $\boldsymbol E$,
\be
c_{c\boldsymbol k}(t) &=&\frac{1}{i\hbar}\int^t_0 dt^\prime
h^*_{vc,\boldsymbol k}(t^\prime) \nonumber \\
&=&
-\frac{1}{i\hbar}\int^t_0 dt^\prime
\frac{i \boldsymbol p_{cv,\boldsymbol K(t^\prime)} \cdot \boldsymbol E(t^\prime) }
{\epsilon_{c, \boldsymbol K(t^\prime)} -\epsilon_{v, \boldsymbol K(t^\prime)}}\frac{e \hbar}{m}
e^{\frac{i}{\hbar}\int^{t^\prime} dt^{\prime \prime} \left \{
\epsilon_{c, \boldsymbol K(t^{\prime\prime})} - \epsilon_{v, \boldsymbol K(t^{\prime\prime})}
\right \}
} \nonumber \\
&=&
-\frac{1}{i\hbar}\int^t_0 dt^\prime
\frac{i \boldsymbol p_{cv,\boldsymbol K(t^\prime)} \cdot \boldsymbol E(t^\prime) }
{\Delta \epsilon_{\boldsymbol K(t^\prime)}}\frac{e \hbar}{m}
e^{\frac{i}{\hbar}\int^{t^\prime} dt^{\prime \prime} \left \{
\Delta \epsilon_{\boldsymbol K(t^{\prime\prime})}
\right \}
},
\label{eq:2band-coeff01}
\ee
where $\Delta \epsilon_{\boldsymbol K(t^{\prime\prime})}$
denotes the energy gap, 
$\epsilon_{c, \boldsymbol K(t^{\prime\prime})} - \epsilon_{v, \boldsymbol K(t^{\prime\prime})}$.

For simplicity, we neglect the time-dependence of the dipole matrix element,
$\boldsymbol p_{cv,\boldsymbol K(t)}/\Delta \epsilon_{\boldsymbol K(t)}$,
and we can simplicity Eq.~(\ref{eq:2band-coeff01}) as
\be
c_{c\boldsymbol k}(t) = -\frac{1}{i\hbar}\int^t_0 dt^\prime
\frac{i \boldsymbol p_{cv,\boldsymbol k} \cdot \boldsymbol E(t^\prime) }
{\Delta \epsilon_{\boldsymbol k}}\frac{e \hbar}{m}
e^{\frac{i}{\hbar}\int^{t^\prime} dt^{\prime \prime} \left \{
\Delta \epsilon_{\boldsymbol K(t^{\prime\prime})}
\right \}
}.
\label{eq:2band-coeff02}
\ee

Furthermore, we expand the contribution from the dynamical phase factor
up to the first order of the electric field, and we obtain
\be
c_{c\boldsymbol k}(t) &=& -\frac{1}{i\hbar}\int^t_0 dt^\prime
\frac{i \boldsymbol p_{cv,\boldsymbol k} \cdot \boldsymbol E(t^\prime) }
{\Delta \epsilon_{\boldsymbol k}}\frac{e \hbar}{m}
e^{\frac{i}{\hbar}\Delta \epsilon_{\boldsymbol k} t^\prime
} \left [ 
1+ \frac{i}{\hbar}\int^{t^\prime}_0 dt^{\prime \prime}
\frac{\hbar \boldsymbol k}{\mu}\cdot \frac{e}{c}\boldsymbol A(t^{\prime\prime})
\right ].
\label{eq:2band-coeff03}
\ee

To proceed with the analysis, 
we assume the following form for the applied electric field,
\be
\boldsymbol E(t) = \boldsymbol e_x E^{P}_x(t)  + \boldsymbol e_y E^{P}_y(t) + 
\boldsymbol e_x E^{DC}_x \Theta(t),
\ee
where $E^P_{x/y}(t)$ is the pump electric field for $x/y$-direction, 
while $E^{DC}_x$ is the static electric field for $x$-direction.
We further assume that the photon energy of the pump pulse is localized 
around the vertical gap $\Delta \epsilon_{\boldsymbol k}$ at $\boldsymbol k$.

For simplicity, here we only consider the excitation on $k_y$-axis, assuming
$k_x=0$. Then, one can evaluate Eq.~(\ref{eq:2band-coeff03}) as
\be
c_{c\boldsymbol k}(t) &=& -\frac{1}{i\hbar}\int^t_0 dt^\prime
\frac{i \boldsymbol p_{cv,\boldsymbol k} \cdot \boldsymbol E^P(t^\prime) }
{\Delta \epsilon_{\boldsymbol k}}\frac{e \hbar}{m}
e^{\frac{i}{\hbar}\Delta \epsilon_{\boldsymbol k} t^\prime
}  \nonumber \\
&&-
\frac{1}{i\hbar}\int^t_0 dt^\prime
\frac{i p_{cv,\boldsymbol k,x} E^{DC}_x }
{\Delta \epsilon_{\boldsymbol k}}\frac{e \hbar}{m}
e^{\frac{i}{\hbar}\Delta \epsilon_{\boldsymbol k} t^\prime
} \frac{i}{\hbar}\int^{t^\prime}_0 dt^{\prime \prime}
\frac{\hbar k_y}{\mu}\cdot \frac{e}{c} A^P_y(t^{\prime\prime}).
\label{eq:2band-coeff04}
\ee

Therefore, the conduction coefficient after infinite time can be evaluated as
\be
c_{c\boldsymbol k}(t\rightarrow \infty) &=& -\frac{1}{i\hbar}
\frac{i \boldsymbol p_{cv,\boldsymbol k} \cdot \boldsymbol {\tilde E}^P(\omega =\Delta \epsilon_{\boldsymbol k}/\hbar) }
{\Delta \epsilon_{\boldsymbol k}}\frac{e \hbar}{m} \nonumber \\
&&-
\frac{1}{i\hbar}
\frac{i p_{cv,\boldsymbol k,x} E^{DC}_x }
{\Delta \epsilon_{\boldsymbol k}}\frac{e \hbar}{m}
\frac{i}{\hbar}\frac{e \hbar k_y}{\mu} 
\frac{\tilde E^P_y(\omega=\Delta \epsilon_{\boldsymbol k}/\hbar)}{\Delta \epsilon^2_{\boldsymbol k}}
\label{eq:2band-coeff05}
\ee
Here, the first term of Eq.~(\ref{eq:2band-coeff05}) corresponds to
the one photon absorption process with the first order perturbation theory,
while the second term corresponds to the nonlinear photocarrier injection
under the presence of the static electric field.

Using the derived coefficients in Eq.~(\ref{eq:2band-coeff05}), 
the conduction population can be expressed as
\be
n_{c\boldsymbol k} &=& \left | c_{c\boldsymbol k}(t\rightarrow \infty) \right |^2 \nonumber \\
&=& \frac{e^2}{m^2} \frac{1}{\Delta \epsilon^2_{\boldsymbol k}}
\left | 
\boldsymbol p_{cv,\boldsymbol k} \cdot \boldsymbol {\tilde E}^P(\omega =\Delta \epsilon_{\boldsymbol k}/\hbar)
\right |^2 \nonumber \\
&& +
i\frac{e^2}{m^2}
\frac{1}{\Delta \epsilon^4_{\boldsymbol k}}
\frac{ek_y}{\mu}p_{cv,\boldsymbol k,x}E^{DC}_x
\left ( 
\boldsymbol p_{cv,\boldsymbol k} \cdot \boldsymbol {\tilde E}^P(\omega =\Delta \epsilon_{\boldsymbol k}/\hbar)
\right )^*\tilde E^P_y(\omega=\Delta \epsilon_{\boldsymbol k}/\hbar) + c.c.
\ee

Assuming perfect circularly-polarized light for the pump, 
$\tilde E^P(\omega) \equiv \tilde E^P_x(\omega) =\pm i \tilde E^P_y(\omega)$,
we evaluate the injected-population difference by the right-handed and the left-handed
circular light as,
\be
\Delta n_{c,\boldsymbol k} = n^{(\circlearrowright)}_{c\boldsymbol k}
-n^{(\circlearrowleft)}_{c\boldsymbol k}
=-4\frac{e^2}{m^2}
\frac{1}{\Delta \epsilon^4_{\boldsymbol k}}
\frac{ek_y}{\mu}E^{DC}_x \left|p_{cv,\boldsymbol k,x}\right|^2
|\tilde E^P(\omega=\Delta \epsilon_{\boldsymbol k}/\hbar)|^2.
\label{eq:2band-pop-imb}
\ee

If we assume the time-reversal symmetry for the ground state Hamiltonian,
the transition momentum holds
\be
\left|p_{cv,\boldsymbol k,x}\right| = \left|p_{cv,-\boldsymbol k,x}\right|. 
\ee
Therefore, Eq.~(\ref{eq:2band-pop-imb}) indicates that
there can be population imbalance in $k$-space under 
circularly-polarized light and static voltage.
Thus, we can conclude that the light-induced Hall current can be
induced even in a topologically trivial (conventional) insulator/semiconductor
with the population imbalance mechanism.

\section{Field strength scaling}

\begin{figure}[htbp]
\centering
\includegraphics[width=0.5\columnwidth]{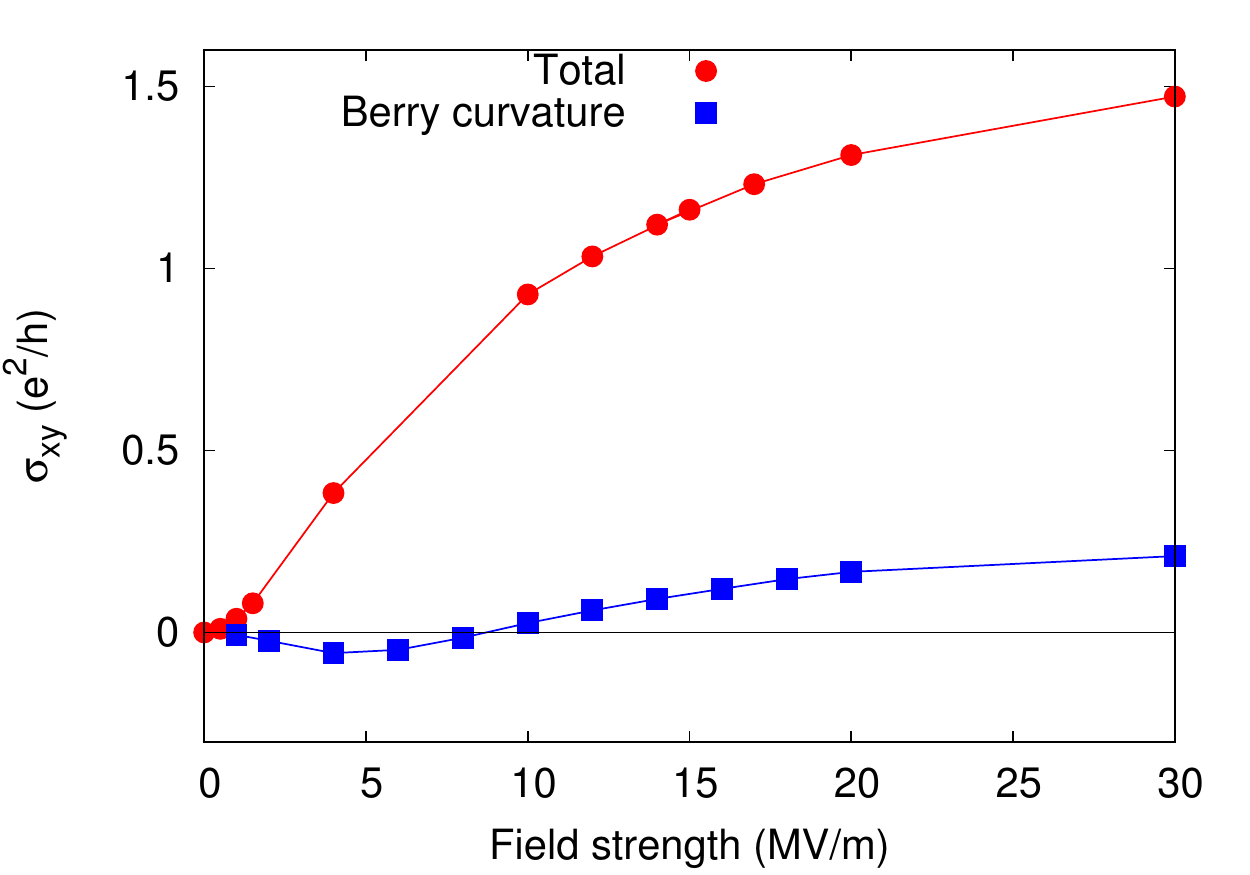}
\caption{\label{fig:intensity_scale}
  The Hall conductivity $\sigma_{xy}$ as a function of applied field strength.
  The total conductivity is shown as the red line, while the topological
  contribution is shown as the blue line.
}
\end{figure}

Here, we investigate the field strength dependence of the light-induced Hall
effect.
Figure~\ref{fig:intensity_scale} shows the computed Hall conductivity
for different field strengths of applied circular laser pulses.
Red line shows the total conductivity, while 
blue line shows the Berry curvature contribution computed by Eq.~(4) in the main text.
As seen from the figure, the total Hall conductivity monotonically increases
with the applied field strength, while the Berry curvature contribution
shows rather complex behaviors with the sign change.
The complex behavior can be understand the high nonlinear population transfer
among different Floquet states because the different Floquet states
have different contribution to the Hall effects, and those contributions
are strongly canceled by each other due to the significant population transfer.
Since the population transfer is highly nonlinear and nontrivial
in the strong field regime,
the resulting Hall conductivity shows the complex behaviors.

\section{Floquet fidelity}

To clarify the relation between the natural orbitals and the Floquet states 
under a continuous-wave laser field,
we introduce \textit{Floquet fidelity}, $S_{\boldsymbol k}$, 
as a measure of the similarity of the two kinds of states.

Under a continuous-wave laser field, the system that obeys 
Eq~(\ref{eq:liouville}) reaches a steady state 
due to the balance of the laser-excitation and the relaxation.
Such steady states can be described
by a time-periodic density matrix,
\be
\rho_{\boldsymbol K(t)}(t) = \rho_{\boldsymbol K(t)}(t+T),
\ee
where $T$ is a optical-cycle of the driving laser field.
Therefore, the natural orbitals $|u^{NO}_{\boldsymbol k}(t)\rangle$, 
which are eigenvectors of the density matrix, 
also have the same time periodicity $T$.

Floquet states $|u^F_{b\boldsymbol k}(t)\rangle$ also have 
the same time periodicity as 
$|u^F_{b\boldsymbol k}(t+T)\rangle=|u^F_{b\boldsymbol k}(t)\rangle$
and they satisfy the time-dependent Schr\"odinger equation~(\ref{eq:tdse}) 
under the periodic driving field as
\be
i\hbar \frac{d}{dt}|\psi^F_{b\boldsymbol k}(t)\rangle
=H_{\boldsymbol K(t)}|\psi^F_{b\boldsymbol k}(t)\rangle
\ee
with 
$|\psi^F_{b\boldsymbol k}(t)\rangle=e^{-i\epsilon^F_{b\boldsymbol k}t/\hbar} |u^F_{b\boldsymbol k}(t)\rangle$, 
where $\epsilon^F_{b\boldsymbol k}$ is so-called Floquet quasienergy.

To define the similarity of the natural orbitals
and the Floquet states, we first consider the cycle-average quantum fidelity
$F_{ij\boldsymbol k}$,
which is equivalent to the squared overlap,
\be
F_{ij\boldsymbol k} =  \frac{1}{T}\int^{T}_{0}dt 
\left |
\langle u^{NO}_{i\boldsymbol k}(t)|u^{F}_{j\boldsymbol k}(t)\rangle\right |^2.
\ee
Then, we define the Floquet fidelity $S_{\boldsymbol k}$ as the absolute value of
the determinant of the Fidelity matrix $F_{\boldsymbol k}$ that
consists of $F_{ij\boldsymbol k}$ as an element,
\be
S_{\boldsymbol k} = \left | \mathrm{det} F_{\boldsymbol k}\right |.
\ee
The Floquet fidelity $S_{\boldsymbol k}$ satisfies
\be
0\le S_{\boldsymbol k} \le 1,
\ee
and $S_{\boldsymbol k}=1$ only if all the natural orbitals
are identical to the Floquet states of the system.
In contrast, if the natural orbitals are fully delocalized
on the Floquet basis, the Floquet fidelity $S_{\boldsymbol k}$
becomes zero.
For example, if the two natural orbitals 
are identical to the two Floquet states, 
\be
u^{NO}_{v\boldsymbol k}(t) &=& u^{F}_{v\boldsymbol k}(t), \\
u^{NO}_{c\boldsymbol k}(t) &=& u^{F}_{c\boldsymbol k}(t),
\ee
the Floquet fidelity becomes one as,
\be
S_{\boldsymbol k}=
\left|
\mathrm{det}
\left(
    \begin{array}{cc}
      1 & 0  \\
      0 & 1
    \end{array}
  \right)\right |=1.
\ee

On the other hand, if the two natural orbitals are fully delocalized
on the Floquet basis such as,
\be
u^{NO}_{v\boldsymbol k}(t) &=& \frac{1}{\sqrt{2}}
\left[ u^{F}_{v\boldsymbol k}(t) + u^{F}_{c\boldsymbol k}(t)\right ], \\
u^{NO}_{c\boldsymbol k}(t) &=& \frac{1}{\sqrt{2}}
\left[ u^{F}_{v\boldsymbol k}(t) - u^{F}_{c\boldsymbol k}(t)\right ],
\ee
the Floquet fidelity becomes zero as
\be
S_{\boldsymbol k}=
\left|
\mathrm{det}
\left(
    \begin{array}{cc}
      1/2 & 1/2  \\
      1/2 & 1/2
    \end{array}
  \right)\right |=0.
\ee

\bibliography{ref}